\documentclass[aps,eqsecnum,preprint,floats,epsf,epsfig,nofootinbib,showpacs]{revtex4}
\usepackage{graphicx}% IncLudE Figure files
\def\be{\begin{eqnarray}}
\def\en{\end{eqnarray}}
\def\non{\nonumber}

\def\bi{\bibitem}
\begin{document}

\title{\Large \bf Analyses of decay constants and light-cone distribution amplitudes
for $s$-wave heavy meson
 }

\author{ \bf  Chien-Wen Hwang\footnote{
t2732@nknucc.nknu.edu.tw}}

\affiliation{\centerline{Department of Physics, National Kaohsiung Normal University,} \\
\centerline{Kaohsiung, Taiwan 824, Republic of China}
 }

%\date{\today}

\begin{abstract}
In this paper, a study of light-cone distribution amplitudes (LCDAs)
for $s$-wave heavy meson are presented in both general and heavy
quark frameworks. Within the light-front approach, the leading twist
light-cone distribution amplitudes, $\phi_M(u)$, and their relevant
decay constants of heavy pseudoscalar and vector mesons, $f_M$, have
simple relations. These relations can be further simplified when the
heavy quark limit is taken into consideration. After fixing the
parameters that appear in both Gaussian and power-law wave
functions, the corresponding decay constants are calculated and
compared with those of other theoretical approaches. The curves and
the first six $\xi$-moments of $\phi_M(u)$ are plotted and
estimated. A conclusion is drawn from these results: Even though the
values of the decay constants of the distinct mesons are almost
equal, the curves of their LCDAs may have quite large differences,
and vice versa. Additionally, in the heavy quark limit, the leading
twist LCDAs, $\Phi_{Qq}(\omega)$ and $\Phi_{Qq}(\omega)$, are
compared with the $B$-meson LCDAs, $\psi_+(\omega)$, suggested by
the other theoretical groups.
\end{abstract}
\maketitle %
%%%%%%%%%%%%%%%%%%%%%%%%%%%%%%%%%%%%%%%%%%%%%%%%%%%%%%%%%%%%%%%%%%%%%%%

\section{Introduction}
The decay constants of heavy mesons with a $c$ or a $b$ quark are
significant quantities and they play an important role in studies of
CP violation, Cabibbo-Kobayashi-Maskawa (CKM) matrix elements, the
$D-\bar D$ or $B-\bar B$ mixing process, and leptonic or nonleptonic
weak decay. Experimentally, new data on the decay constants of the
pseudoscalar mesons $f_D$ and $f_{D_s}$ have been reported
\cite{CLEO08,CLEO09,Belle08,CLEO09a} which has provided a precise
method for comparing different theoretical calculations and for
checking their accuracy. During the last decade, the decay constants
of both pseudoscalar and vector heavy mesons have been studied by
lattice simulations \cite{Bec}, the relativistic quark model
\cite{Choi,CJBc,BS1,BS2,Ebert}, and the field correlator method
\cite{FC}. The light-cone distribution amplitudes (LCDAs) of hadrons
are key ingredients in the description of the various exclusive
processes of quantum chromodynamics (QCD), and their role is
analogous to that of parton distributions in inclusive processes. In
terms of the Bethe-Salpeter wave functions,
$\varphi(u_i,k_{i\perp})$, the LCDAs, $\phi(u_i)$, are defined by
retaining the momentum fractions, $u_i$, and integrating out the
transverse momenta, $k_{i\perp}$ \cite{LB}. They provide essential
information on the nonperturbative structure of the hadron for the
QCD treatment of exclusive reactions and they play a central role in
all known factorization formulas. Specifically, the leading twist
LCDAs describe the probability amplitudes for finding the hadron in
a Fock state with the minimum number of constituents.
Experimentally, the fact that $B$-physics exclusive processes are
under investigation in BaBar and Belle experiments also urges the
detailed study of
hadronic LCDAs. In the literature, %the LCDAs of light mesons have been
%studied by many non-perturbative approaches, such as the QCD sum
%rules \cite{CZ,Bakulev,Ball1,Yang1}, lattice calculation
%\cite{Ali,Braun}, chiral quark model from the instanton vacuum
%\cite{Petrov,Nam}, Nabmbu-Jona-Lasinio model
%\cite{Praszalowicz,Arriola}, and the light-front quark model
%\cite{Ji2,Ji1}.
%s for the heavy mesons, there are two
the LCDAs of heavy quarkonia have been estimated by various
nonperturbative approaches, such as QCD sum rules
\cite{Braguta1,Braguta2,Braguta3,Braguta4}, NRQCD factorization
\cite{MS}, and the light-front quark model \cite{Ji3,me1,me2}. As
for heavy-light mesons, the LCDAs of $B$-meson $\psi_{\pm}$ were
first introduced within the heavy quark effective theory (HQET)
\cite{GN}, and the following studies were intensive
\cite{japan1,LN,BIK,HWZ,MW,GW,KMO,CL,HQW,LN1,GW1,YOR,BF,KT,GO},
whereas the ones of other heavy-light mesons were discussed in a
non-HQET framework \cite{Ji4}.

In the past decade, the most significant progress made in the QCD
description of hadronic physics was, perhaps, in the avenue of heavy
quark dynamics. The analysis of heavy hadron structures has been
tremendously simplified by the heavy quark symmetry (HQS) proposed
by Isgur and Wise \cite{IW1,IW2} and HQET developed from QCD in
terms of $1/m_Q$ expansion \cite{Georgi,EH1,EH2}. HQET has provided
a systematic framework for studying symmetry breaking $1/m_Q$
corrections (for a review, see \cite{Neubert}). Moreover, in terms
of heavy quark expansion, HQET offered a new framework for the
systematic study of the inclusive decays of heavy mesons
\cite{CGG,Bigi,MW1,Mannel}. However, the general properties of heavy
hadrons, namely their decay constants, transition form factors and
structure functions etc., are still incalculable within QCD, even in
the infinite quark mass limit with the utilization of HQS and HQET.
%Indeed, HQS cannot provide a direct description of nonperturbative
%QCD dynamics to other heavy hadron properties.
Hence, although HQS and HQET have simplified heavy quark dynamics, a
complete first-principles QCD description of heavy hadrons has still
been lacking due to the unknown nonperturbative QCD dynamics.

This paper has focused on the study of the decay constants and the
leading twist LCDAs of pseudoscalar ($D$, $D_s$, $B$, $B_s$, $B_c$)
and vector ($D^*$, $D_s^*$, $B^*$, $B^*_s$, $B^* _c$) mesons
within both general and heavy quark frameworks. %The motivation
%of this study is as follows. Since the discoveries of $J/\psi$ and
%$\Upsilon$, occurring more than thirty years ago, a great deal of
%information on heavy quarkonium levels and their transitions has
%been accumulated \cite{PDG08}. The numerous transitions between
%heavy quarkonium states are classified as strong and radiative
%decays, which shed light on aspects of QCD in both the perturbative
%and the non-perturbative regimes (for a recent review see
%\cite{EGMR}). In particular, some experimental results regarding
%$\chi_{cJ}$ mesons have recently been reported
%\cite{CLEO1,CLEO2,CLEO3,CLEO4}.
From the definitions of the decay constant and LCDA (or quark
distribution amplitude (DA)) \cite{LB}, these two properties seemed
to be closely related. In terms of a detailed analysis, the purpose
of this study is to transparently realize the relation between the
decay constant and LCDA of the heavy meson. We believe that a
thorough understanding of these universal nonperturbative objects
would be of great benefit when analyzing the hard exclusive
processes with heavy meson production or annihilation. Additionally,
%it is known that heavy-light meson is relevant to the relativistic
%treatments.
%Although non-relativistic QCD (NRQCD) is a powerful theoretical tool
%for separating high-energy modes from low-energy contributions, in
%most cases the calculation of low-energy hadronic matrix elements
%has relied on model-dependent non-perturbative methods.
in this study, the $s$-wave heavy meson has been explored within the
light-front quark model (LFQM), which is a promising analytic method
for solving the nonperturbative problem of hadron physics
\cite{BPP}, as well as offering many insights into the internal
structures of bound states. The basic ingredient in LFQM is the
relativistic hadron wave function which generalizes distribution
amplitudes by including transverse momentum distributions, and which
contains all the information of a hadron from its constituents. The
hadronic quantities are represented by the overlap of wave functions
and can be derived in principle. The light-front wave function is
manifestly a Lorentz invariant, expressed in terms of internal
momentum fraction variables which are independent of the total
hadron momentum. Moreover, the fully relativistic treatment of quark
spins and center-of-mass motion can be carried out using the
so-called Melosh rotation \cite{LFQM}. This treatment has been
successfully applied to calculate phenomenologically many important
meson decay constants and hadronic form factors \cite{Jaus1,Jaus12,
CCH1, Jaus2, CCH2, Hwang}. Therefore, the main purpose of this study
was the calculation of the leading twist LCDAs of $s$-wave heavy
mesons within LFQM.

The remainder of this paper is organized as follows. In Sec. II, the
leading twist LCDAs of $s$-wave heavy meson states are derived
within general and heavy quark frameworks. In Sec. III, the
formulations of LFQM within the general and heavy quark frameworks
are reviewed briefly, and the decay constants and the leading twist
LCDAs then extracted. In Sec. IV, numerical results and discussions
are presented. Finally, the conclusions are given in Sec. V.

%%%%%%%%%%%%%%%%%%%%%%%%%%%%%%%%%%%%%%%%%%%%%%%%%%%%%%%%%%%%%%%%%%%%%%%%%
\section{Leading twist LCDAs of $s$-wave mesons}
%%%%%%%%%%%%%%%%%%%%%%%%%%%%%%%%%%%%%%%%%%%%%%%%%%%%%%%%%%%%%%%%%%%%
\subsection{General Framework}
The amplitudes of the hard processes involving $s$-wave mesons can
be described by the matrix elements of gauge-invariant nonlocal
operators, which are sandwiched between the vacuum and the meson
states,
 \be
 \langle 0 |\bar q (x) \Gamma [x,-x] q(-x) | H(P,\epsilon)\rangle,
 \label{nonlocal}
 \en
where $P$ is the meson momentum, $\epsilon$ is the polarization
vector (of course, $\epsilon$ does not exist in the case of
pseudoscalar meson), $\Gamma$ is a generic notation for the Dirac
matrix structure, and the path-ordered gauge factor is:
 \be
 [x,y]={\textrm{P exp}}\left[ig_s\int^1_0 dt(x-y)_\mu
 A^\mu(tx+(1-t)y)\right].
 \en
This factor is equal to unity in the light-cone gauge which is
equivalent to the fixed-point gauge, $(x-y)_\mu A^\mu (x-y)=0$, as
the quark-antiquark pair is at the lightlike separation
\cite{Yang2}. For simplicity, the gauge factor will not be shown
below.

The asymptotic expansion of exclusive amplitudes in powers of large
momentum transfer is governed by the expanding amplitude, Eq.
(\ref{nonlocal}), shown in powers of deviation from the light-cone
$x^2=0$. The two lightlike vectors, $p$ and $z$, can be introduced
by
 \be
 p^2=0,~~~~~~z^2=0,
 \en
so that $p \to P$ in the limit $M_H^2 \to 0$ and $z \to x$ for $x^2
= 0$. From this it follows that \cite{Ball1}
 \be
 z^\mu &=& x^\mu - P^\mu \frac{1}{M_H^2} \left[Px-\sqrt{(Px)^2-x^2
 M_H^2}\right] \non \\
 &=& x^\mu - P^\mu \frac{x^2}{2 P z}+ O(x^4), \non \\
  p^\mu &=& P^\mu -z^\mu \frac{M^2_H}{2 P z},\label{Pp}
 \en
where $P x \equiv P \cdot x$ and $P z=p z=\sqrt{(P x)^2-x^2 M_H^2}$.
In addition, if the meson is assumed to move in a positive
$\hat{e}_3$ direction, then $p^+$ and $z^-$ are the only nonzero
components of $p$ and $z$, respectively, in an infinite momentum
frame. For the vector meson, the polarization vector $\epsilon^\mu$
is decomposed into longitudinal and transverse projections as
 \be
 \epsilon^\mu_{\|}=\frac{\epsilon z}{p z}\left(p^\mu-z^\mu\frac{M^2_H}{2p z}\right),
 ~~~\epsilon^\mu_{\perp}=\epsilon^\mu-\epsilon^\mu_{\|}, \label{epsilon}
 \en
respectively.

LCDAs are defined in terms of the matrix element of the nonlocal
operator in Eq. (\ref{nonlocal}). For the pseudoscalar $(P)$ and
vector $(V)$ mesons, LCDAs can be defined as
 \be
 \langle 0|\bar q (z) \gamma^\mu\gamma_5 q (-z)|P(P)\rangle &=& i f_P \int^1_0 du~e^{i\xi p z}\left[p^\mu
  \phi_P(u)+z^\mu \frac{M_P^2}{2p z}g_P(u)\right], \label{P}\\
 \langle 0|\bar q (z) \gamma^\mu q (-z)|V(P,\epsilon_{\lambda=0})\rangle
 &=&f_V M_V \int^1_0 du~e^{i\xi p z}\Big\{p^\mu \frac{\epsilon z}{p z}
 \phi_{V\|}
 (u) +\epsilon^\mu_{\perp}g_{V\perp}(u)\non \\
 &&~~~~~~~~~~~~~~~~~~~~~~- z^\mu \frac{\epsilon z}{2 (p z)^2}M^2_V
 g_{V3}(u)\Big\}, \label{VL}\\
 \langle 0|\bar q (z) \sigma^{\mu\nu} q (-z)|V(P,\epsilon_{\lambda=\pm 1})\rangle
 &=&f^\perp_V \int^1_0 du~e^{i\xi p z}\Big\{(\epsilon^\mu_{\perp} p^\nu-\epsilon^\nu_{\perp} p^\mu)
 \phi_{V_\perp} (u)\non \\
 &+&(p^\mu z^\nu-p^\nu z^\mu)\frac{M^2_V \epsilon z}{(p z)^2} h_{V\|}(u)
 +(\epsilon^\mu_{\perp} z^\nu-\epsilon^\nu_{\perp} z^\mu) \frac{M^2_V}{2 p z}
 h_{V3}(u)\Big\},\label{VT}
 \en
where $u$ is the momentum fraction and $\xi \equiv (1-u)-u =1- 2 u$.
Here $\phi_P$, $\phi_{V\|}$, and $\phi_{V\perp}$ are the leading
twist-$2$ LCDAs, and the others contain contributions from
higher-twist operators. The leading twist LCDAs are normalized as
 \be
 \int^1_0 du \phi (u) = 1, \label{normal}
 \en
and can be expanded \cite{CZ} in Gegenbauer polynomials
$C^{3/2}_n(\xi)$ as
 \be
 \phi(\xi,\mu)= \phi_{\rm as}(\xi)\left[\sum_{l=0}^\infty
 a_l(\mu)C_l^{3/2}(\xi)\right].
 \en
where $\phi_{\rm as}(\xi)=3(1-\xi^2)/4$ is the asymptotic quark
distribution amplitude and $a_l(\mu)$ are the Gegenbauer moments
which describe to what degree the quark distribution amplitude
deviates from the asymptotic one. $C^{3/2}_l(\xi)$'s have the
orthogonality integrals
 \be
 \int^1_{-1} (1-\xi^2)C^{3/2}_l(\xi) C^{3/2}_m (\xi) d\xi =
 \frac{2(l+1)(l+2)}{2l+3}~\delta_{lm}. \label{orthogonal}
 \en
Then $a_l$ can be obtained by using the above orthogonality
integrals as
 \be
 a_l(\mu) &=&\frac{2(2
 l+3)}{3(l+1)(l+2)}\int^1_{-1}C^{3/2}_l(\xi)\phi(\xi,\mu)d\xi.
 \en
An alternative approach to parameterize the quark distribution
amplitude is to calculate the so-called $\xi$-moments:
 \be\label{ximoments}
 \langle \xi^n\rangle_\mu=\int^1_{-1} d\xi~\xi^n \phi(\xi,\mu),
 \en
as calculated in this work.

 To disentangle the twist-$2$ LCDAs from higher twists in Eqs.
(\ref{P}) $\sim$ (\ref{VT}), the twist-$2$ contribution of the
relevant nonlocal operator $\bar q (z) \Gamma  q(-z)$ must be
derived. For the $\Gamma = \gamma^\mu (\gamma_5)$ case, the leading
twist-$2$ contribution contains contributions of the operators which
are fully symmetric in the Lorentz indices \cite{Ball2,BB}:
 \be
 [\bar q (-z)\gamma^\mu (\gamma_5)
 q(z)]_2=\sum^\infty_{n=0} \frac{1}{n!}\bar q
 (0)\bigg\{\frac{(z\cdot \widehat{D})^n}{n+1} \gamma^\mu +
 \frac{n (z\cdot \widehat{D})^{n-1}}{n+1}\widehat{D}^\mu {\not
 \!z}\bigg\}(\gamma_5)q(0), \label{twist2expand}
 \en
where $\widehat{D}=\overrightarrow{D}-\overleftarrow{D}$ and
$\overrightarrow{D}=\overrightarrow{\partial}-ig B^a (\lambda^a/2)$.
The sum can be expressed in terms of a nonlocal operator,
 \be
 [\bar q (-z)\gamma^\mu (\gamma_5)
 q(z)]_2= \int^1_0 dt \frac{\partial}{\partial z_\mu} \bar q (-t z) \not
 \!z (\gamma_5) q(t z). \label{t1}
 \en
Taking the matrix element between the vacuum and the $s$-wave meson
state, we obtained:
 \be
 \langle 0 |[\bar q (-z)\gamma^\mu \gamma_5
 q(z)]_2|P(P)\rangle &=& i f_P \int^1_0 du
 \phi_P(u) \Bigg\{p^\mu e^{i\xi p z} +(P^\mu -p^\mu)\int^1_0 dte^{i\xi
 t p z}\Bigg\}, \label{Pphi}\\
 \langle 0 |[\bar q (-z)\gamma^\mu
 q(z)]_2|V(P,\epsilon_{\lambda=0})\rangle &=& f_V M_V \int^1_0 du
 \phi_{V\|}(u) \Bigg\{p^\mu \frac{\epsilon z}{p z} e^{i\xi p z}\non
 \\
 &&~~~~~~~~~~~~~~~~~~+\left(\epsilon^\mu-p^\mu \frac{\epsilon z}{p z}\right)\int^1_0 dte^{i\xi
 t p z}\Bigg\},\label{Vphi}
 \en
The derivations of (\ref{Vphi}) as shown in Ref. \cite{Ball2}, are
applied to those of Eq. (\ref{Pphi}). We use Eq.
(\ref{twist2expand}), and then expand the right-hand sides of Eqs.
(\ref{Pphi}) and (\ref{Vphi}) as
 \be
 &&~~~\sum^\infty_{n=0} \frac{1}{n!}\langle 0|\bar q
 (0)\bigg\{\frac{(z\cdot \widehat{D})^n}{n+1} \gamma^\mu +
 \frac{n (z\cdot \widehat{D})^{n-1}}{n+1}\widehat{D}^\mu {\not
 \!z}\bigg\}\gamma_5 q(0)|P(P)\rangle\non \\
 &=& i f_P \sum^\infty_{n=0}
 \frac{i^n}{n!}\int^1_0 du \phi_P(u) (\xi p z)^n \Bigg\{p^\mu +(P^\mu -p^\mu)\int^1_0
 dt t^n\Bigg\},\label{Pn}\\
 &&~~~\sum^\infty_{n=0} \frac{1}{n!}\langle 0|\bar q
 (0)\bigg\{\frac{(z\cdot \widehat{D})^n}{n+1} \gamma^\mu +
 \frac{n (z\cdot \widehat{D})^{n-1}}{n+1}\widehat{D}^\mu {\not
 \!z}\bigg\}q(0)|V(P,\epsilon)\rangle\non \\
 &=& f_V M_V \sum^\infty_{n=0}
 \frac{i^n}{n!}\int^1_0 du \phi_{V\|}(u) (\xi p z)^n \Bigg\{p^\mu \frac{\epsilon z}{p z}
  +\left(\epsilon^\mu -p^\mu \frac{\epsilon z}{p z}\right)\int^1_0
 dt t^n\Bigg\},\label{Vn} \en
respectively. Picking $n=0$ in Eqs. (\ref{Pn}) and (\ref{Vn}), we
obtain
 \be
 \langle 0|\bar q (0) \gamma^\mu \gamma_5 q(0)|P(P)\rangle &=& i f_P P^\mu
 \int^1_0 du \phi_P (u), \label{P0}\\
 \langle 0|\bar q (0) \gamma^\mu q(0)|V(P),\epsilon_{\lambda=0}\rangle &=& f_V M_V \epsilon^\mu
 \int^1_0 du \phi_{V\|} (u).\label{V0}
 \en
From the normalization of Eq. (\ref{normal}), we have
 \be
 \langle 0|\bar q
 \gamma^\mu \gamma_5 q|P(P)\rangle &=& i f_P P^\mu, \non \\
 \langle 0|\bar q
 \gamma^\mu q|V(P,\epsilon)\rangle &=& f_V M_V \epsilon^\mu
 \en
which are taken as the definitions of decay constants $f_P$ and
$f_V$ in the literature.

Next, we consider the case of $\Gamma = \sigma_{\mu\nu}$, where the
leading twist-$2$ contribution contains contributions of the
operators:
 \be
 [\bar q (-z)\sigma^{\mu\nu}
 q(z)]_2&=&\sum^\infty_{n=0} \frac{1}{n!}\bar q
 (0)\bigg\{\frac{(z\cdot \widehat{D})^n}{2 n+1} \sigma^{\mu\nu} +
 \frac{n (z\cdot \widehat{D})^{n-1}}{2 n+1}\widehat{D}^\mu
 \sigma^{\bullet\nu}\non \\
 &&\qquad\qquad\quad+\frac{n (z\cdot \widehat{D})^{n-1}}{2
 n+1}\widehat{D}^\nu
 \sigma^{\mu\bullet}\bigg\}q(0). \label{twist2expandsigma}
 \en
The sum can also be represented in terms of nonlocal operators:
 \be
 [\bar q (-z)\sigma^{\mu\nu}  q(z)]_2= \int^1_0 dt \left[\frac{\partial}
 {\partial z_\mu} \bar q (-t^2 z) \sigma^{\bullet\nu}  q(t^2 z)+z_\alpha \frac{\partial}
 {\partial z_\nu} \bar q (-t^2 z) \sigma^{\mu\alpha}  q(t^2 z)\right]. \label{t2}
 \en
Returning to Eq. (\ref{VT}), it can be rewritten as
 \be
 \langle 0|\bar q (z) \sigma_{\mu\nu} q (-z)|V(P,\epsilon_{\lambda=\pm 1})\rangle
 &=&f^\perp_V \int^1_0 du~e^{i\xi p z}\Big\{(\epsilon_\mu P_\nu-\epsilon_\nu P_\mu)
 \phi_{V\perp} (u)\non \\
 &+&(p_\mu z_\nu-p_\nu z_\mu)\frac{M^2_V \epsilon z}{(p
 z)^2}[h_{V\|}(u)-\phi_{V\perp} (u)]\non \\
 &+&(\epsilon_{\perp\mu} z_\nu-\epsilon_{\perp\nu} z_\mu) \frac{M^2_V}{2 p z}
 [h_{V3}(u)-\phi_{V\perp} (u)]\Big\}.\label{b1}
 \en
We sandwich both sides of Eq. (\ref{t2}) between the vacuum and the
vector meson state as
 \be
 &&\langle 0|[\bar q (z) \sigma_{\mu\nu} q(-z)]_2
 |V(P,\epsilon_{\lambda=\pm 1})\rangle\non \\
 &=& \int^1_0 dt \Bigg[\frac{\partial}
 {\partial z^\mu} \langle 0|\bar q (-t^2 z) \sigma_{\bullet\nu} q(t^2
 z)|V(P,\epsilon)\rangle +z^\alpha \frac{\partial}
 {\partial z^\nu} \langle 0|\bar q (-t^2 z) \sigma_{\mu\alpha} q(t^2 z)
 |V(P,\epsilon)\rangle\Bigg] \non \\
 &=& f_V^\perp \int^1_0 du \Bigg\{\phi_{V\perp}(u)\Bigg[(\epsilon_\mu P_\nu
 -\epsilon_\nu P_\mu)\int^1_0 dt e^{i\xi t^2 p z}+2 p z {\cal S_{\mu\nu}}
 (i\xi)\int^1_0 dt t^2 e^{i\xi t^2 p z}\Bigg]\non \\
 &&\qquad\qquad+\big(h_{V\|}(u)-\phi_{V\perp}(u)\big)\Bigg[{\cal U_{\mu\nu}}\int^1_0dte^{i\xi t^2 p z}+
 2 p z {\cal T_{\mu\nu}}(i\xi)\int^1_0 dt t^2e^{i\xi t^2 p
 z}\Bigg]\bigg\}. \label{b3}
 \en
The integral is performed as
 \be
 i \xi \int^1_0 dt t^2 e^{i\xi t^2 p z} = \frac{1}{2 p
 z} \int^1_0 dt t \frac{\partial}{\partial t}e^{i\xi t^2 p z}
 =\frac{1}{2 p z} \left[e^{i\xi p z}-\int^1_0 d t e^{i\xi t^2 p
 z}\right], \label{b4}
 \en
and then we substitute Eq. (\ref{b3}) for Eq. (\ref{b4}) to obtain
 \be
 &&\langle 0 |[\bar q (-z)\sigma^{\mu\nu}
 q(z)]_2|V(P,\epsilon_{\lambda=\pm 1})\rangle \non \\
 &=& i f^\perp_V \int^1_0 du
 \Bigg\{\phi_{V\perp}(u) \bigg[{\cal S^{\mu\nu}}  e^{i\xi p z}
 +\bigg((\epsilon^\mu P^\nu-\epsilon^\nu P^\mu)-
 {\cal S^{\mu\nu}}\bigg)\int^1_0 dte^{i\xi t^2 p z}\bigg]\non \\
 &&\quad\quad\qquad+\bigg(h_{V\|}(u)-\phi_{V\perp}(u)\bigg)
 \Bigg[{\cal T^{\mu\nu}} e^{i\xi p z}+\bigg({\cal U^{\mu\nu}}-
 {\cal T^{\mu\nu}}\bigg)\int^1_0 dte^{i\xi t^2 p
 z}\Bigg]\Bigg\},\label{Vphisigma}
 \en
where
 \be
 {\cal S^{\mu\nu}}&=&\frac{1}{2}\Bigg[(\epsilon^\mu P^\nu-\epsilon^\nu
 P^\mu)
 -(\epsilon^{\mu}_\perp z^\nu -\epsilon^{\nu}_\perp z^\mu)\frac{M_V^2}{2 p z}\Bigg],\non \\
 {\cal T^{\mu\nu}}&=&\frac{\epsilon z M_V^2}{2 (p z)^2} (p^\mu z^\nu-p^\nu z^\mu)
 ,\qquad\qquad
 {\cal U^{\mu\nu}}=\frac{M_V^2}{p z} (\epsilon^\mu z^\nu-\epsilon^\nu
 z^\mu).
 \en
%The derivations of Eq. (\ref{Vphisigma}) are shown in Appendix A.
In contrast to Eqs. (\ref{Pphi}) and (\ref{Vphi}), the twist-$2$
LCDAs do not disentangle entirely from the higher twists in Eq.
(\ref{Vphisigma}). Taking the product with $\epsilon_{\perp\mu}
z_\nu$ in Eq. (\ref{Vphisigma}) to obtain
 \be
 \langle 0 |[\bar q (-z)\sigma^{\mu\bullet}\epsilon_{\perp\mu}
  \gamma_5 q(z)]_2|V(P,\epsilon_{\lambda=\pm 1})\rangle &=& i f^\perp_V \int^1_0 du
 \phi_{V\perp}(u) \frac{1}{2}(\epsilon\cdot \epsilon_\perp P z)\bigg[e^{i\xi p z}
 +\int^1_0 dte^{i\xi t^2 p z}\bigg],\non \\ \label{Vphisigmadis}
 \en
we then use Eq. (\ref{twist2expandsigma}) and expand the right-hand
side of Eq. (\ref{Vphisigmadis}) as
 \be
 &&~~~\sum^\infty_{n=0} \frac{1}{n!}\langle 0|\bar q
 (0)\frac{(n+1)(z\cdot \widehat{D})^n}{2 n+1}
 \sigma^{\mu\bullet}\epsilon_{\perp\mu} \gamma_5
 q(0)|V(P,\epsilon_{\lambda=\pm 1})\rangle \non \\
 &=& f^\perp_V  \sum^\infty_{n=0}
 \frac{i^n}{n!}\int^1_0 du \phi_{V\perp}(u) \frac{1}{2}(\epsilon\cdot \epsilon_\perp P z)
 (\xi p z)^n \Bigg[1+\int^1_0 dt t^{2n}\Bigg].\label{Vnsigma}
 \en
Picking $n=0$ in Eq. (\ref{Vnsigma}), we obtain
 \be
 \langle 0|\bar q(0)\sigma^{\mu\bullet}\epsilon_{\perp\mu}
 q(0)|V(P,\epsilon_{\lambda=\pm 1})\rangle &=& f^\perp_V\int^1_0 du \phi_{V\perp}(u) (\epsilon\cdot
 \epsilon_\perp P z).\label{Vnsigmafm}
 \en
From the normalization of Eq. (\ref{normal}), we have
 \be
 \langle 0|\bar q(0)\sigma^{\mu\bullet}\epsilon_{\perp\mu}
 q(0)|V(P,\epsilon_{\lambda=\pm 1})\rangle &=& f^\perp_V (\epsilon\cdot
 \epsilon_\perp P z),\label{Vnsigmafm1}
 \en
which is consistent with the usual definition of $f^\perp_V$ as
 \be
 \langle 0|\bar q(0)\sigma^{\mu\nu}
 q(0)|V(P,\epsilon_{\lambda=\pm 1})\rangle &=& f^\perp_V
 (\epsilon^\mu P^\nu-\epsilon^\nu P^\mu),\label{Vnsigmafmdef}
 \en
%It is worth noting that the author of Ref. \cite{Yang2} also
%considered an approach that disentangled the twist-$2$ LCDAs from
%the higher twists, in the case of an axial vector meson state
%($\Gamma = \sigma^{\mu\nu}\gamma_5 $): Besides $z_\nu$, Eq.
%(\ref{Vphisigma}) has taken the product with a term proportional to
%$(\epsilon_\mu P z-P_\mu \epsilon z )$. We find this approach
%equivalent to ours. The derivation is as follows. The term
%$(\epsilon_\mu P z-P_\mu \epsilon z )$ can be expanded by using Eqs.
%(\ref{Pp}) and (\ref{epsilon}) as
% \be
% \epsilon_\mu P z-P_\mu \epsilon z &=& p_\mu \frac{e z}{p z}P z
% -z_\mu \frac{e z M^2_A}{2 (p z)^2} P z + \epsilon_{\perp\mu} P z
% -p_\mu e z -z_\mu\frac{e z M^2_A}{2 p z}\non \\
% &=&  -z_\mu \frac{e z M^2_A}{p z}+\epsilon_{\perp\mu} P z.
% \en
%The first term of last line has no contribution to the result
%because $\sigma^{\mu\nu}$ is antisymmetric. %Therefore the
%disentangled approach used in Ref. \cite{Yang2} is equivalent to
%ours.
%%%%%%%%%%%%%%%%%%%%%%%%%%%%%%%%%%%%%%%%%%%%%%%%%%%%%%%%%%%%%%%%%%%%%%%%%%%%%%%%%
\subsection{Heavy Quark Framework}
%%%%%%%%%%%%%%%%%%%%%%%%%%%%%%%%%%%%%%%%%%%%%%%%%%%%%%%%%%%%%%%%%%%%%%%%%%%%%%%%%
%In the past decade, the most significant progress made in the QCD
%description of hadronic physics is perhaps in the avenue of heavy
%quark dynamics. The analysis of heavy hadron structures has been
%tremendously simplified by heavy quark symmetry (HQS) proposed by
%Isgur and Wise \cite{IW1,IW2} and the heavy quark effective theory
%(HQET) developed from QCD in terms of $1/m_Q$ expansion
%\cite{Georgi,EH1,EH2}.
In general, the theoretical description of meson properties relies
on the bound state models with a relativistic normalization:
 \be
 \langle M(P')|M(P)\rangle=2P^0 (2\pi)^3
 \delta^3(P'-P).\label{normnorm}
 \en
At low energies, however, these models have little connection to the
fundamental theory of QCD. Then the reliable predictions are often
made based on symmetries. A well-known example is HQS
\cite{Neubert}, which arises since the Compton wave-length, $1/m_Q$,
of a heavy quark bound inside a hadron is much smaller than a
typical hadronic distance (about $1$ fm), and $m_Q$ is unimportant
for the low energy properties of the state. For a heavy-light meson
system, it is more natural to use velocity $v^\mu$ instead of
momentum variables. Then it is appropriate to work with a mass
independent normalization of a heavy-light meson state:
 \be
 \langle \widehat{M}(v')|\widehat{M}(v)\rangle=2v^0 (2\pi)^3
 \delta^3(\bar{\Lambda}
 v'-\bar{\Lambda}v), \label{normHQ}
 \en
where $\bar \Lambda=M-m_Q$ is the so-called residual center mass of
a heavy-light meson. The relation between these two bound states is
 \be
 |M(P)\rangle =\sqrt{M}|\widehat{M}(v)\rangle. \label{states}
 \en
In addition, the heavy quark field can be expanded as \cite{Neubert}
 \be
 Q(x)=e^{-im_Q v\cdot x}\left[1+\frac{1}{iv\cdot D+2 m_Q-i\varepsilon}
 i\not\!\! D_\perp\right]h_v(x),
 \en
where $h^*_v(x)$ is a field describing a heavy antiquark with
velocity $v$. Then the current $\bar q \Gamma Q$ can be represented
as
 \be
 \bar q \Gamma Q =\bar q \Gamma \left(1+\frac{i\not\!\! D_\perp}{2 m_Q}
 +\cdot\cdot\cdot\right)h_v. \label{currentE}
 \en
Substituting Eqs. (\ref{states}) and (\ref{currentE}) into the
definitions of LCDAs, Eqs. (\ref{P}) $\sim$ (\ref{VT}) give
 \be
 \langle 0|\bar q (z) \gamma^\mu\gamma_5 h_v (-z)|\widehat{P}(v)\rangle &=& i F_P
 \int^\infty_0 d\omega~e^{i\omega v z}\left[v^\mu
 \Phi_P(\omega)+z^\mu \frac{1}{2v z}G_P(\omega)\right], \label{Pv}\\
 \langle 0|\bar q (z) \gamma^\mu h_v (-z)|\widehat{V}(v,\epsilon_{\lambda=0})\rangle
 &=&F_V \int^\infty_0 d\omega~e^{i\omega v z}\Big\{v^\mu \frac{\epsilon z}{v z}
 \Phi_{V\|}
 (\omega) +\epsilon^\mu_{\perp}G_{V\perp}(\omega)\non \\
 &&~~~~~~~~~~~~~~~~~~~~~~- z^\mu \frac{\epsilon z}{2 (v z)^2}
 G_{V3}(\omega)\Big\}, \label{VLv}\\
 \langle 0|\bar q (z) \sigma^{\mu\nu} h_v (-z)|\widehat{V}(v,\epsilon_{\lambda=\pm 1})\rangle
 &=&F^\perp_V \int^\infty_0 d\omega~e^{i\omega v z}\Big\{(\epsilon^\mu_{\perp} v^\nu-
 \epsilon^\nu_{\perp} v^\mu)
 \Phi_{V\perp} (\omega)\non \\
 &+&(v^\mu z^\nu-v^\nu z^\mu)\frac{\epsilon z}{(v z)^2} H_{V\|}(\omega)
 +(\epsilon^\mu_{\perp} z^\nu-\epsilon^\nu_{\perp} z^\mu) \frac{1}{2 v z}
 H_{V3}(\omega)\Big\},\label{VTv}
 \en
where $F_M=\sqrt{M}f_M$, $\Phi_i (\omega)=\phi_i(u)/M$, and $\omega$
was first introduced in Ref. \cite{CZL} as the product of
longitudinal momentum fraction $u$ of the light (anti)quark and the
mass of heavy meson $M$, namely $\omega=u M$. Following a similar
process, the leading twist LCDAs are obtained as
 \be
 \langle 0|\bar q (0) \gamma^\mu \gamma_5 h_v(0)|\widehat{P}(v)\rangle &=& i F_P v^\mu
 \int^\infty_0 d\omega \Phi_P (\omega), \label{P0v}\\
 \langle 0|\bar q (0) \gamma^\mu h_v(0)|\widehat{V}(v),\epsilon_{\lambda=0}
 \rangle &=& F_V \epsilon^\mu
 \int^\infty_0 d\omega \Phi_{V\|} (\omega), \label{V0v}\\
 \langle 0|\bar q(0)\sigma^{\mu\bullet}\epsilon_{\perp\mu}
 h_v(0)|\widehat{V}(v,\epsilon_{\lambda=\pm 1})\rangle &=& F^\perp_V(\epsilon\cdot
 \epsilon_\perp v z) \int^\infty_0
 d\omega \Phi_{V\perp}(\omega).\label{Vpv}
 \en

The authors of Ref. \cite{GN} defined two quark-antiquark wave
functions in momentum space $\psi_{\pm} (\omega)$ of a heavy-light
meson in terms of the matrix element:
 \be
 \langle 0 |\bar q(z) \Gamma h_v(-z)|\widehat{M}
 (v)\rangle=f\int^\infty_0 e^{i\omega v z} d\omega {\rm
 Tr}\left\{\left[\psi_+(\omega)+\frac{\not\!z}{2vz}[\psi_-(\omega)-
 \psi_+(\omega)]\right]{\cal M}(v)\Gamma\right\},
 \en
where $f=F_M /2$ and
 \be
 {\cal M}(v) = \frac{1+\not\! v}{2}\Bigg\{\begin{array}{l}
            -i\gamma_5, {\rm~~~~for~pseudoscalar~meson~} M(v), \\
            \not\!\epsilon, {\rm~~~~~~~~~for~vector~meson~}
            M^*(v, \epsilon).
            \end{array}
 \en
Evaluating the trace for various choices of $\Gamma$ and taking the
heavy quark limit, they obtained
 \be \label{PVV}
 \Phi_P(\omega)=\Phi_{V\|}(\omega)=\Phi_{V\perp}(\omega)=\psi_+(\omega),
 \en
and the normalization conditions
 \be
 \int^\infty_0 d\omega \psi_+(\omega)=1.\label{normw}
 \en
In addition, the authors of Ref. \cite{GN} defined the moments of
$\psi_+(\omega)$ as
 \be \label{momentw}
 \langle \omega^n \rangle_+ =\int^\infty_0 d\omega
 \psi_+(\omega)\omega^n,
 \en
and used the equations of light and heavy quarks to obtain the
relation between the first moment and the residual center mass:
 \be\label{omegaLambda}
 \langle \omega \rangle_+ = \frac{4}{3}\bar{\Lambda}.
 \en
%%%%%%%%%%%%%%%%%%%%%%%%%%%%%%%%%%%%%%%%%%%%%%%%%%%%%%%%%%%%%%%%%%%
\section{Formulism in Light-Front Approach }
%%%%%%%%%%%%%%%%%%%%%%%%%%%%%%%%%%%%%%%%%%%%%%%%%%%%%%%%%%%%%%%%%%%
\subsection{General Framework}

An $s$-wave meson bound state, consisting of a quark, $q_1$, and an
antiquark, $\bar q_2$, with total momentum $P$ and spin $J$, can be
written as (see, for example \cite{CCH1})
 \be
 |M(P, S, S_z)\rangle =\int &&\{d^3k_1\}\{d^3k_2\} ~2(2\pi)^3
 \delta^3(\tilde P -\tilde k_1-\tilde k_2)~\non\\
 &&\times \sum_{\lambda_1,\lambda_2}
 \Psi^{SS_z}(\tilde k_1,\tilde k_2,\lambda_1,\lambda_2)~
 |q_1(k_1,\lambda_1) \bar q_2(k_2,\lambda_2)\rangle,\label{lfmbs}
 \en
where $k_1$ and $k_2$ are the on-mass-shell light-front momenta,
 \be
 \tilde k=(k^+, k_\bot)~, \quad k_\bot = (k^1, k^2)~,
 \quad k^- = \frac{m_q^2+k_\bot^2}{k^+},
 \en
and
 \be
 &&\{d^3k\} \equiv \frac{dk^+d^2k_\bot}{2(2\pi)^3}, \nonumber \\
 &&|q(k_1,\lambda_1)\bar q(k_2,\lambda_2)\rangle
 = b^\dagger(k_1,\lambda_1)d^\dagger(k_2,\lambda_2)|0\rangle,\\
 &&\{b(k',\lambda'),b^\dagger(k,\lambda)\} =
 \{d(k',\lambda'),d^\dagger(k,\lambda)\} =
 2(2\pi)^3~\delta^3(\tilde k'-\tilde k)~\delta_{\lambda'\lambda}.
 \nonumber
 \en
In terms of the light-front relative momentum variables $(u,
\kappa_\bot)$ defined by
 \be
 && k^+_1=(1-u) P^{+}, \quad k^+_2=u P^{+}, \nonumber \\
 && k_{1\bot}=(1-u) P_\bot+\kappa_\bot, \quad k_{2\bot}=u
 P_\bot-\kappa_\bot,
 \en
the momentum-space wave function $\Psi^{SS_z}$ can be expressed as
 \be
 \Psi^{SS_z}(\tilde k_1,\tilde k_2,\lambda_1,\lambda_2)
 = \frac{1}{\sqrt N_c}
 R^{SS_z}_{\lambda_1\lambda_2}(u,\kappa_\bot)~ \varphi(u,
 \kappa_\bot),\label{Psi}
 \en
where $\varphi(u,\kappa_\bot)$ describes the momentum distribution
of the constituent quarks in the bound state, and
$R^{SS_z}_{\lambda_1\lambda_2}$ constructs a state of definite spin
($S,S_z$) out of the light-front helicity ($\lambda_1,\lambda_2$)
eigenstates. Explicitly,
 \be
 R^{SS_z}_{\lambda_1 \lambda_2}(u,\kappa_\bot)
 =\sum_{s_1,s_2} \langle \lambda_1|
  {\cal R}_M^\dagger(1-u,\kappa_\bot, m_1)|s_1\rangle
 \langle \lambda_2|{\cal R}_M^\dagger(u,-\kappa_\bot, m_2)
 |s_2\rangle \langle \frac{1}{2}\,\frac{1}{2};s_1
 s_2|\frac{1}{2}\frac{1}{2};SS_z\rangle,
 \en
where $|s_i\rangle$ are the usual Pauli spinors and ${\cal R}_M$ is
the Melosh transformation operator~\cite{Jaus1}:
 \be
 \langle s|{\cal R}_M (u_i,\kappa_\bot,m_i)|\lambda\rangle
 %&=&\frac{\bar
 %u_D(k_i,s) u(k_i,\lambda)}{2 m_i}=-\frac{\bar
 %v_D(k_i,s) v(k_i,\lambda)}{2 m_i}
 %\non\\
 &=&\frac{m_i+u_i M_0
 +i\vec \sigma_{s\lambda}\cdot\vec \kappa_\bot \times
 \vec n}{\sqrt{(m_i+u_i M_0)^2 + \kappa^{2}_\bot}},
 \en
with $u_1=1-u$, $u_2=u$, and %$u_{(D)}$, a Dirac spinor in light-front form,
$\vec n =(0,0,1)$ a unit vector in the $\hat {z}$-direction. In
addition,
 \be
 M_0^2&=&(e_1+e_2)^2=\frac{m_1^2+\kappa^2_\bot}{1-u}+\frac{m_2^2+\kappa^2_\bot}{u},
 \\
 e_i&=&\sqrt{m^2_i+\kappa^2_\perp+\kappa^2_z}.\non
 \en
where $\kappa_z$ is the relative momentum in $\hat{z}$ direction and
can be written as
 \be \label{eq:Mpz}
  \kappa_z=\frac{u M_0}{2}-\frac{m^2_2+\kappa^2_\perp}{2 u M_0}.
 \en
$M_0$ is the invariant mass of $q\bar q$ and generally different
from mass $M$ of the meson which satisfies $M^2=P^2$. This is due to
the fact that the meson, quark and antiquark cannot be
simultaneously on-shell. We normalize the meson state as
 \be
 \langle M(P',S',S'_z)|M(P,S,S_z)\rangle = 2(2\pi)^3 P^+
 \delta^3(\tilde P'- \tilde P)\delta_{S'S}\delta_{S'_z S_z}~,
 \label{wavenor}
 \en
in order that
 \be
 \int \frac{du\,d^2\kappa_\bot}{2(2\pi)^3}~
 |\varphi(u,\kappa_\bot)|^2=1.
 \label{momnor}
 \en
In general, for any function ${\cal F}(|\vec{\kappa}|)$,
$\varphi(u,\kappa_\perp)$ has the form of
 \be
 \varphi(u,\kappa_\perp)=N \sqrt{\frac{d\kappa_z}{du}}{\cal F}(|\vec{\kappa}|),\label{F}
 \en
where normalization factor $N$ is determined from
Eq.~(\ref{momnor}).

%Note that $u_D(p,s)=u(p,\lambda) \langle \lambda|{\cal
%R}^\dagger_M|s\rangle$ and, consequently, under rotation a state
%$|q(p,\lambda)\rangle \langle \lambda|{\cal R}^\dagger_M|s\rangle$
%transforms like $|q(p,s)\rangle$, i.e. its transformation does not
%depend on its momentum.

In practice, it is more convenient to use the covariant form of
$R^{SS_z}_{\lambda_1\lambda_2}$ \cite{Jaus1,CCH2,cheung}:
 \be
 R^{SS_z}_{\lambda_1\lambda_2}(u,\kappa_\bot)
 =\frac{\sqrt{k_1^+ k_2^+}}{\sqrt2~{\widetilde M_0}(M_0+m_1+m_2)}
 \bar u(k_1,\lambda_1)(\not\!\!\bar P+M_0)\Gamma
 v(k_2,\lambda_2), \label{covariantp}
 \en
where
 \be
 &&\widetilde M_0\equiv\sqrt{M_0^2-(m_1-m_2)^2},\qquad\quad \bar
 P\equiv k_1+k_2,\non \\
 &&\bar u(k,\lambda) u(k,\lambda')=\frac{2
 m}{k^+}\delta_{\lambda,\lambda'},\qquad\quad \sum_\lambda u(k,\lambda)
 \bar u(k,\lambda)=\frac{\not\!k +m}{k^+},\non \\
 &&\bar v(k,\lambda) v(k,\lambda')=-\frac{2
 m}{k^+}\delta_{\lambda,\lambda'},\qquad\quad \sum_\lambda v(k,\lambda)
 \bar v(k,\lambda)=\frac{\not\!k -m}{k^+}.
 \en
For the pseudoscalar and vector mesons, we have:
 \be
 \Gamma_{P}=\gamma_5,\quad
 \Gamma_{V}=-\not\!\epsilon(\lambda),\label{Gamma}
 \en
where
 \be
 &&\epsilon^\mu_{\lambda=\pm 1} =
 \left[\frac{2}{ P^+} \vec \epsilon_\bot (\pm 1) \cdot
 \vec P_\bot,\,0,\,\vec \epsilon_\bot (\pm 1)\right],\non \\
 &&\vec \epsilon_\bot (\pm 1)=\mp(1,\pm i)/\sqrt{2}, \non\\
 &&\epsilon^\mu_{\lambda=0}=\frac{1}{M_0}\left(\frac{-M_0^2+P_\bot^2}{
 P^+},P^+,P_\bot\right).   \label{polcom}
 \en
Equations (\ref{covariantp}) and (\ref{Gamma}) can be further
reduced by the applications of equations of motion on the spinors
\cite{CCH2}:
 \be
 R^{SS_z}_{\lambda_1\lambda_2}(u,\kappa_\bot)
 =\frac{\sqrt{k_1^+ k_2^+}}{\sqrt2~{\widetilde M_0}}
 \bar u(k_1,\lambda_1)\Gamma' v(k_2,\lambda_2), \label{covariantfurther}
 \en
where
 \be
 \Gamma'_{P}=\gamma_5,\quad
 \Gamma'_{V}=-\not\!\epsilon+\frac{\epsilon\cdot(k_1-k_2)}
 {M_0+m_1+m_2}.\label{Gammap}
 \en
%The derivations of Eqs. (\ref{covariantfurther}) and (\ref{Gammap})
%are shown in Appendix C.

%\subsection{Analysis of Leading twist LCDAs}
Next, the matrix elements of Eqs. (\ref{P0}), (\ref{V0}), and
(\ref{Vnsigmafm}) are calculated within LFQM, and the relevant
leading twist LCDAs are extracted. For the pseudoscalar meson state,
we substitute Eqs. (\ref{lfmbs}), (\ref{Psi}), and
(\ref{covariantfurther}) into Eq. (\ref{P0}) to obtain
 \be
 \langle 0|\bar q_2 \gamma^\mu \gamma_5 q_1|P(P)\rangle&=&N_c\int\{d^3
 k_1\}\sum_{\lambda_1,\lambda_2}\Psi^{SS_z}(k_1,k_2,\lambda_1,\lambda_2)
 \langle 0 |\bar q_2 \gamma^\mu \gamma_5 q_1|q_1\bar q_2\rangle\non \\
 &=& i\sqrt{N_c}\int\{d^3 k_1\}\frac{\sqrt{k_1^+ k_2^+}}{\sqrt2~{\widetilde
 M_0}}\varphi{\rm Tr}\bigg[\gamma^\mu \gamma_5  \Bigg(\frac{\not\!k_1+m_1}{k_1^+}\Bigg)\gamma_5
 \Bigg(\frac{\not\!k_2-m_2}{k^+_2}\Bigg)\bigg]\non \\
 &=&i f_P P^\mu \int du \phi(u).
 \en
For the "good" component, $\mu=+$, the leading twist LCDA, $\phi_P$,
is extracted as
 \be
 \phi_P(u) = \frac{2\sqrt{6}}{f_P}\int \frac{d^2
 \kappa_\perp}{2(2\pi)^3}\frac{[(1-u) m_2+u m_1]}{\sqrt{u(1-u)}\widetilde M_0}
 \varphi(u,\kappa_\perp).\label{Su}
 \en
A similar process is used for the vector meson which corresponded to
Eqs. (\ref{V0}) and (\ref{Vnsigmafm}), and then the leading twist
LCDAs are extracted as
 \be
 \phi_{V\|} (u)&=&\frac{2\sqrt{6}}{f_V}\int \frac{d^2
 \kappa_\perp}{2(2\pi)^3}\frac{\varphi(u,\kappa_\perp)}{\sqrt{u(1-u)}\widetilde M_0}
 \Bigg\{u m_1+(1-u) m_2
 +\frac{2\kappa_\perp^2}{M_0+m_1+m_2} \Bigg\},\label{V}\\
 \phi_{V\perp} (u)&=&\frac{2\sqrt{6}}{f^\perp_V}\int \frac{d^2
 \kappa_\perp}{2(2\pi)^3}\frac{\varphi(u,\kappa_\perp)}{\sqrt{u(1-u)}\widetilde M_0}
 \Bigg\{u m_1+(1-u) m_2+\frac{\kappa_\perp^2}{M_0+m_1+m_2}\Bigg\}.\label{Vk}
 \en
From the normalization of Eq. (\ref{normal}), we found not only that
the equations of $f_P$ and $f_V$ were consistent with that of
\cite{Choi}, but also that the decay constants and the leading twist
LCDAs has the simple relations
 \be
 f_P+f_V=2f_V^\perp,\qquad \phi_P(u)+\phi_{V\|}(u)=2\phi_{V\perp}(u).
 \en
%Furthermore, if we take $m_1 \to \infty$, that is, the heavy quark
%limit in the heavy-light meson, then an inequality $m_1 \simeq M_0
%\gg m_2$ is obtained. From Eqs. (\ref{Su}) $\sim$ (\ref{Vk}), the
%decay constants and the leading twist LCDAs can be simplified as:
% \be
% f_P\simeq f_V \simeq f^\perp_V,\qquad \phi_P\simeq \phi_{V\|}\simeq
% \phi_{V\perp},
% \en
%which are independent of the form of $F(|\vec{\kappa}|)$. These are
%consistent with the so-called heavy quark symmetry (HQS) within the
%$s$-wave heavy-light mesons.
%%%%%%%%%%%%%%%%%%%%%%%%%%%%%%%%%%%%%%%%%%%%%%%%%%%%%%%%%%%%%%%%%%%%%%%%%%%%
\subsection{Heavy Quark Framework}
%%%%%%%%%%%%%%%%%%%%%%%%%%%%%%%%%%%%%%%%%%%%%%%%%%%%%%%%%%%%%%%%%%%%%%%%%%%%
If one takes $m_1 =m_Q \to \infty$, that is, the heavy quark limit
in the heavy-light meson, then two inequalities, $m_Q \simeq M_0 \gg
m_2$, $\kappa_\perp$ and $u \to 0$, are obtained. From Eqs.
(\ref{Su}) $\sim$ (\ref{Vk}), the decay constants and the leading
twist LCDAs are simplified as
 \be
 f_P\simeq f_V \simeq f^\perp_V \propto F_M,
 \qquad \phi_P\simeq \phi_{V\|}\simeq
 \phi_{V\perp} \propto \Phi_M,
 \en
which are independent of the form of ${\cal F}(|\vec{\kappa}|)$.
These are consistent with HQS between the $s$-wave heavy-light
mesons. The exact form of $\Phi_M$, however, must be derived by the
redefinition of the meson bound state. Let us consider the bound
states of heavy mesons in the heavy quark limit:
 \begin{eqnarray}  \label{hqslfb}
  |\widehat{M}(v;S,S_z)\rangle &=& \int \{d^3q\}\{d^3k_2\} 2(2\pi)^3 \delta^3(
    \overline{\Lambda}\tilde{v}-\tilde{q}-\tilde{k}_2) \nonumber \\
    &\times& \sum_{\lambda_1,\lambda_2}
    \widehat{\Psi}^{SS_z} (\omega,\kappa_{\bot},\lambda_1,\lambda_2)
    b_v^\dagger(q, \lambda_1) d^\dagger (k_2, \lambda_2)|0\rangle,
 \end{eqnarray}
where $q=k_1-m_Q v$ is the residual momentum of heavy quark. The
operators $b_v^\dagger(q,\lambda_1)$ create a heavy quark with
\begin{eqnarray}
    \{ b_v (q,\lambda_1), ~ b_{v'}^\dagger (q',\lambda'_1) \}
        &=&2 (2\pi)^3 \delta_{vv'}\delta^3(\tilde{q}-\tilde{q}')
        \delta_{\lambda_1 \lambda'_1},
\end{eqnarray}
The relative transverse and longitudinal momenta, $\kappa_\bot$ and
$\kappa_z$, are obtained by
 \begin{equation}
 \kappa_\bot=k_{2\bot}-\omega v_\bot,~~\kappa_z
 ={\omega\over{2}}-{m_2^2+\kappa_\bot^2\over{2 \omega}}.\label{kzH}
 \end{equation}
The momentum-space wave-function $\widehat{\Psi}^{SS_z}$ can be
expressed as
 \be
 \widehat{\Psi}^{SS_z}(\omega,\kappa_{\bot},\lambda_1,\lambda_2)
 = \frac{1}{\sqrt N_c}
 \widehat{R}^{SS_z}_{\lambda_1\lambda_2}(\omega,\kappa_\bot)~
 \widehat{\varphi}^{SS_z}(\omega, \kappa_\bot),\label{hatPsi}
 \en
where
 \begin{equation}  \label{spin}
  \widehat{R}^{SS_z}(\omega, \kappa_{\bot}, \lambda_1, \lambda_2)
    =\frac{k_2^+}{\sqrt{2}
        \sqrt{(\omega+m_2)^2+\kappa^2_\perp}} ~ \bar u (v,\lambda_1)
        \Gamma v(k_2,\lambda_2)
 \end{equation}
with $\Gamma=\gamma_5 (-\not\!\hat{\epsilon})$ for $S=0(1)$,
 \be
 &&\hat{\epsilon}^\mu_{\lambda=\pm 1} =
 \left[\frac{2}{ v^+} \vec \epsilon_\bot (\pm 1) \cdot
 \vec v_\bot,\,0,\,\vec \epsilon_\bot (\pm 1)\right],\non\\
 &&\hat{\epsilon}^\mu_{\lambda=0}=\left(\frac{-1+v_\bot^2}{
 v^+},v^+,v_\bot\right),  \label{Hpolcom}
 \en
and $u(v,\lambda_1)$ is the spinor for the heavy quark,
 \begin{equation}
    \sum_\lambda u(v,\lambda)\overline{u}(v,\lambda)
        = \frac{{\not \! v}+1}{v^+}.
\end{equation}
The normalization of the heavy meson bound states can then be given
by
 \begin{equation}  \label{nmc2}
    \langle \widehat{M}(v',S',S'_z) |\widehat{M}(v,S,S_z)\rangle = 2(2\pi)^3 v^+
        \delta^3(\overline{\Lambda}v'-\overline{\Lambda}v)
        \delta_{SS'} \delta_{S_zS'_z},
 \end{equation}
which not only leads to %two things: first, the heavy meson bound state
%$|\widehat{M}(v;S,S_z)\rangle$ in this model rescales the one
%$|M(P;S,S_z)\rangle$ in the LFQM by
%$|M\rangle=\sqrt{M}|\widehat{M}\rangle$
Eq. (\ref{states}), but also to the space part
$\widehat{\varphi}^{SS_z}(\omega,\kappa_\bot^2)$ (called the
light-front wave function) in Eq.(\ref{hqslfb}) which has the
following wave function normalization condition:
 \begin{equation} \label{nwf}
   \int \frac{d\omega d^2\kappa_\bot} {2(2\pi)^3}
        |\widehat{\varphi}^{SS_z} (\omega, \kappa^2_\bot)|^2 = 1.
 \end{equation}
In principle, the heavy quark dynamics are completely described by
HQET, which is given by the $1/m_Q$ expansion of the heavy quark QCD
Lagrangian:
\begin{equation}
    {\cal L} = \overline{Q} (i \not \! \! D - m_Q) Q \nonumber
    = \sum_{n=0}^\infty \Bigg({1 \over 2m_Q}
        \Bigg)^n {\cal L}_n.
        \label{hqcdl}
\end{equation}
Therefore, $|\widehat{M}(v;S,S_z)\rangle$ and
$\widehat{\varphi}^{SS_z} (\omega,\kappa^2_{\bot})$ are then
determined by the leading Lagrangian ${\cal L}_0=\bar h_v iv\cdot D
h_v$. The authors of Ref. \cite{HCCZ} have shown, from the
light-front bound state equation, that $\widehat{\varphi}^{SS_z}
(U,\kappa^2_{\bot})$ must be degenerate for $S=0$ and $S=1$. As a
result, we can simply write
\begin{equation} \label{msiwf}
    \widehat{\varphi}^{SS_z}(\omega,\kappa^2_\bot) = \widehat{\varphi}(\omega,\kappa^2_\bot)
\end{equation}
in the heavy quark limit. Equation (\ref{hqslfb}) together with
Eqs.~(\ref{spin}) and (\ref{msiwf}) are then the heavy meson
light-front bound states in the heavy quark limit that obeyed HQS.
From the normalization conditions of Eqs. (\ref{momnor}) and
(\ref{nwf}), we obtain the relation between the wave functions
$\varphi(u,\kappa^2_\perp)$ and
$\widehat{\varphi}(\omega,\kappa^2_\perp)$:
 \be \label{scale}
 \varphi(u,\kappa^2_\perp)=\sqrt{M}\widehat{\varphi}(\omega,\kappa^2_\perp).
 \en
%Thus Eqs. (\ref{P0}), (\ref{V0}), and (\ref{Vnsigmafm}) can be
%rewritten as:
% \be
% \langle 0|\bar q (0) \gamma^\mu \gamma_5 h_v(0)|\widehat{P}(v)\rangle &=& i F_P v^\mu
% \int^\infty_0 d\omega \Phi_P (\omega), \label{PH0}\\
% \langle 0|\bar q (0) \gamma^\mu h_v(0)|\widehat{V}(v,\hat{\epsilon}_{\lambda=0})\rangle &=& F_V  \hat{\epsilon}^\mu
% \int^\infty_0 d\omega \Phi_{V\|} (\omega),\label{VH0}\\
% \langle 0|\bar q(0)\sigma^{\mu\bullet}\hat{\epsilon}_{\perp\mu}
% h_v(0)|\widehat{V}(v,\hat{\epsilon}_{\lambda=\pm 1})\rangle &=& F^\perp_V(\hat{\epsilon}\cdot
% \hat{\epsilon}_\perp v z)\int^\infty_0 d\omega \Phi_{V\perp}(\omega),\label{VnsigmafmH}
% \en
%where the leading twist LCDAs in HQS are normalized as Eq.
%(\ref{normw}).
 %\be
 %\int^\infty_0 d\omega \Phi(U)=1.
 %\en
Next, the matrix elements of Eqs. (\ref{P0v}), (\ref{V0v}), and
(\ref{Vpv}) can be calculated, and the relevant leading twist LCDAs
extracted. For the pseudoscalar meson state, we substitute Eqs.
(\ref{hqslfb}), (\ref{hatPsi}), and (\ref{spin}) into Eq.
(\ref{P0v}) to obtain
 \be
 \langle 0|\bar q \gamma^\mu \gamma_5 h_v|P(v)\rangle&=&N_c\int\{d^3
 k_2\}\sum_{\lambda_1,\lambda_2}\widehat{\Psi}^{SS_z}(\omega,\kappa_\perp,\lambda_1,\lambda_2)
 \langle 0 |\bar q_2 \gamma^\mu \gamma_5 h_v|Q\bar q_2\rangle\non \\
 &=& i\sqrt{N_c}\int\{d^3 k_2\}\frac{k_2^+}{\sqrt2~{\sqrt{(\omega+m_2)^2+\kappa^2_\perp}
 }}\widehat{\varphi}{\rm Tr}\bigg[\gamma^\mu\gamma_5 \Bigg(\frac{\not\!v+1}{v^+}\Bigg)\gamma_5
 \Bigg(\frac{\not\!k_2-m_2}{k^+_2}\Bigg)\bigg]\non \\
 &=&i F_P v^\mu \int d\omega \Phi_P(\omega).
 \en
For the "$+$" component, the leading twist LCDA $\Phi_P$ is
extracted as
 \be
 \Phi_P(\omega) = \frac{2\sqrt{6}}{F_P}\int \frac{d^2
 \kappa_\perp}{2(2\pi)^3}\frac{\omega+ m_2}{\sqrt{(\omega+m_2)^2+\kappa^2_\perp}}
 \widehat{\varphi}(\omega,\kappa_\perp).\label{PHIP}
 \en
In contrast with $\phi(u)$, % which means the distribution of the
%longitudinal momentum fraction carried by the light degree of
%freedom
$\Phi(\omega)$ represents the distribution of the longitudinal
momentum carried by the light degree of freedom. A similar process
is used for the vector meson which corresponds to Eqs. (\ref{V0v})
and (\ref{Vpv}), and the results are
 \be
 F_P=F_V=F^\perp_V,\qquad \Phi_P (\omega)=\Phi_{V\|} (\omega)=\Phi_{V\perp}
 (\omega),
 \en
which are consistent with Eqs. (\ref{PVV}) and (\ref{normw}).
%with $F_M=\sqrt{M}f_P$ and
% \be
% \Phi_M (U) = \frac{2\sqrt{6}}{F_M}\int \frac{d^2 \kappa_\perp}{2
% (2\pi)^3}
% \frac{U+m_2}{\sqrt{(U+m_2)^2+\kappa^2_\perp}}\hat{\varphi}(U,\kappa^2_\perp).
% \en
%%%%%%%%%%%%%%%%%%%%%%%%%%%%%%%%%%%%%%%%%%%%%%%%%%%%%%%%%%%%%%%%%%%%%%%%%%%
\section{Numerical results and discussions}
In this section, the decay constants and LCDAs of $D^{(*)}$,
$D_s^{(*)}$, $B^{(*)}$, $B_s^{(*)}$, and $B_c^{(*)}$ are studied. We
consider two kinds of ${\cal F}(|\vec \kappa|)$, one is the Gaussian
type, the other is the power-law type:
%\cite{fpHwang}:
 \be\label{FF}
 {\cal F}^g(|\vec \kappa|)&=&{\rm exp}\bigg(-\frac{|\vec \kappa|^2}{2
 \beta^2}\bigg), \\
 {\cal F}^p(|\vec \kappa|)&=&\bigg(\frac{1}{1+|\vec \kappa|^2/
 \beta^2}\bigg)^2,
 \en
then the corresponding wave functions are %and $\hat{\varphi}$ are
 \be
 \varphi^g(u,\kappa_\perp)&=&4\bigg(\frac{\pi}
 {\beta^2}\bigg)^{3/4}\sqrt{\frac{e_1 e_2}{u (1-u) M_0}}~{\rm
 exp}\bigg[-\frac{\kappa_\perp^2+(\frac{u M_0}{2}-\frac{m^2_2+\kappa^2_\perp}{2 u M_0})^2}{2
 \beta^2}\bigg],\label{Gaussian1s}\\
 \varphi^p(u,\kappa_\perp)&=&8\bigg(\frac{2\pi}
 {\beta^3}\bigg)^{1/2}\sqrt{\frac{e_1 e_2}{u (1-u) M_0}}~\Bigg[\frac{\beta^2}
 {\kappa_\perp^2+(\frac{u M_0}{2}-\frac{m^2_2+\kappa^2_\perp}{2 u M_0})^2+\beta^2}\Bigg]^2,\label{Powerlaw1s}
 %\hat{\varphi}(U,\kappa_\perp)&=&4\bigg(\frac{\pi}
 %{\beta^2}\bigg)^{3/4}\sqrt{\frac{1}{2}+\frac{m^2_2+\kappa^2_\perp}{2
 %U^2}}~{\rm exp}
 %\bigg(-\frac{\kappa_\perp^2+(\frac{U}{2}
 %-\frac{m^2_2+\kappa^2_\perp}{2 U})^2}{2\beta^2}\bigg),\label{Gaussian1sH}
 \en
and can be used to calculate decay constant $f$, the %and $F_M$, the
LCDAs $\phi(u)$, %and $\Phi_M(U)$,
and the $\xi$-moments of $\phi(u)$. Prior to the numerical
calculations, the parameters $m_1$, $m_2$ and $\beta$, which
appeared in the wave function, have to first determined. For the
light quark masses, we used the %values $m_{u,d}\equiv m_q=0.251$ GeV
%and $m_s=0.445 \pm0.036$ GeV, which were fitted for the
decay constants $f_\pi$, $f_K$ and the mean square radii $\langle
r^2_{\pi^+}\rangle$, $\langle r^2_{K^0}\rangle$ to fit
$m_{u,d}(\equiv m_q)$ and $m_s$ \cite{fpHwang}. For the heavy quark
masses, however, the relevant measurements are insufficient. We %used
%the heavy quark masses $m_c=1.38$ GeV and $m_b=4.78$ GeV which were
determined $m_c$ and $m_b$ by the mass of the spin-weighted average
of the %triplet $p$-wave
heavy quarkonium states and its variational principle for the
relevant Hamiltonian \cite{me2}.

As regards parameter $\beta$, it is determined by the decay constant
of the heavy meson. Recently the CLEO collaboration updated their
data concerning ${\rm Br}(D^+ \to \mu^+\nu)$ and an average value
was reported \cite{RS10}: $f_{D^+}=206.0 \pm 8.9$ MeV. In addition,
the authors of Ref. \cite{RS} averaged ${\rm Br}(B^- \to \tau^-\bar
\nu)$ from the Belle \cite{Belle} and Babar \cite{Babar1,Babar2}
collaborations and extracted $f_B = 204 \pm 31$ MeV. The parameters
$\beta_{cq}$ and $\beta_{bq}$ can then be determined. As mentioned
in the previous work \cite{fpHwang}, the ratios,
$\beta_{cs}/\beta_{cq}$ and $\beta_{bs}/\beta_{bq}$ can be related
to the $SU(3)$ symmetry breaking, that is, $m_s/m_q$ as follows:
 \be
 \frac{\Delta M_{D_sD_s^*}}{\Delta M_{DD^*}}=\frac{m_q}{m_s}
 \left(\frac{\beta_{cs}}{\beta_{cq}}\right)^3, \qquad
 \frac{\Delta M_{B_sB_s^*}}{\Delta M_{BB^*}}=\frac{m_q}{m_s}
 \left(\frac{\beta_{bs}}{\beta_{bq}}\right)^3. \label{splitting}
 \en
Therefore, $\beta_{cs}$ and $\beta_{bs}$ are not independent
parameters. Concerning the decay constants of $B_c$, %the heavy-heavy meson,
%we use $f_{J/\psi}=416\pm 7$ MeV and $f_\Upsilon=715\pm 5$ MeV,
%which are obtained from the experimental data $\Gamma(J/\psi
%(\Upsilon )\to e^+e^-)$, and
we quote the average result of QCD sum rules \cite{QCDsumrules}:
$f_{B_c}=360$ MeV to extract the parameter $\beta_{bc}$. All the
parameters are listed in Table 1.
\begin{table}[ht!]
\caption{\label{tab:parameters} Input values of quark masses and
$\beta$'s (MeV). }
%\begin{ruledtabular}
 {\footnotesize
\begin{tabular}{|c|c|c|c|c|c|c|c|c|c|}\hline
 &$m_q$ & $m_s$ & $m_c$ & $m_b$ & $\beta_{cq}$  & $\beta_{cs}$ &
 $\beta_{bq}$ & $\beta_{bs}$ & $\beta_{bc}$  \\ \hline
 $\varphi^g$&$251$ & $445\pm 36$ & $1380$ & $4780$ & $465\pm 22$ & $567\pm 42$ &
   $587\pm 74$ & $727\pm 120$ & $815$ \\
 $\varphi^p$&$172$ & $296\pm 12$ & $1360$ & $4770$ & $505\pm 25$ & $608\pm 40$ &
   $575\pm 77$ & $706\pm 113$ & $815$ \\  \hline
\end{tabular}}
%\end{ruledtabular}
\end{table}

Next, we used the parameters in Table 1 as input to calculate the
decay constants $f_P$, $f_V$, and $f_V^\perp$ of the relevant heavy
mesons. The values of the ratios $f_V/f_P, f_{P'}/f_P, f_{V'}/f_V$
are also included. Tables 2 and 3 show a comparison of the results
of this work with other theoretical calculations.
 \begin{table}%[ht!]
 \caption{\label{tab:decayconstant} Decay constants of the
pseudoscalar and vector heavy mesons (MeV). Linear and HO are the
different potentials in Refs. \cite{CJBc,Choi}, FC is the field
correlators, BS is the Bethe-Salpeter equation, and RQM is the
relativistic quark model.
 } %The ratio $f_V/f_P$ is
%also included.
%\begin{ruledtabular}
 {\footnotesize
 \begin{tabular}{|c|c|c|c|c|c|c|c|}\hline
 & Experiment & This work\footnote{The value is obtained by $\varphi^g(\varphi^p)$.} & Linear(HO) & FC\cite{FC}  & BS\cite{BS1,BS2} & Lattice\cite{Bec}
 & RQM\cite{Ebert}\\\hline
 $f_D$ &$206.0\pm 8.9$ \cite{RS10} &\underline{$206.0\pm8.9$}(\underline{$206.0\pm8.9$}) & $211(194)$ & $210\pm10$ & $230\pm 25$ & $211\pm14^{+0}_{-12}$ &$234$  \\
 $f_{D^*}$& & $259.6\pm 14.6$($306.3^{+18.2}_{-17.7}$) & $254(228)$& $273\pm13$ &$340\pm 23$ & $245\pm20^{+0}_{-2}$& $310$\\
 $f_{D^*}^\perp$ & & $232.7\pm 11.7$($256.2^{+13.6}_{-13.3}$) &  &  & &  & \\
 $f_{D_s}$ &$260.7\pm 6.5$\cite{RS10} & $267.4\pm 17.9$($259.7\pm13.7$) & $248(233)$ & $260\pm 10$ & $248\pm 27$ & $231\pm12^{+6}_{-0}$& $268$  \\
 $f_{D^*_s}$ && $338.7\pm 29.7$($391.0\pm 28.9$)  & $290(268)$ & $307\pm 18$ & $375\pm 24$ & $272\pm12^{+0}_{-20}$& $315$\\
 $f_{D^*_s}^\perp$ && $303.1\pm 23.8$ ($325.3\pm 21.5$) &  & &  & & \\
 %$f_{\eta_c}$ & $335\pm 75$\cite{CLEO01}& $354\pm 5$ & $326(354)$ &  & $292\pm 25$ & &   \\
 %$f_{J/\psi}$ &$416\pm 7$\cite{PDG08}& \underline{$416\pm 7$}  & $360(395)$ &  & $459\pm 28$ & & \\
 %$f_{J/\psi}^\perp$ && $385\pm 6$  &  & &  & & \\
 $f_B$ & $204\pm31$\footnote{This value is extracted by the branching ratio:
 ${\cal B} (B^-\to \tau^- \bar \nu)=(1.42 \pm 0.43)\times
10^{-4}$ \cite{RS}.} &\underline{$204\pm 31$}(\underline{$204\pm 31$}) & $189(180)$ & $182\pm 8$ & $196\pm29$ & $179\pm18^{+26}_{-9}$ &$189$  \\
 $f_{B^*}$& & $225\pm 38$($249^{+44}_{-42}$) & $204(193)$& $200\pm 10$ &$238\pm 18$ & $196\pm24^{+31}_{-2}$& $219$\\
 $f_{B^*}^\perp$ & & $214\pm 34$($226\pm 37$) &  &  & &  & \\
 $f_{B_s}$ && $281\pm 54$($270\pm 47$) & $234(237)$ & $216\pm 8$  & $216\pm 32$ & $204\pm16^{+28}_{-0}$& $218$ \\
 $f_{B^*_s}$ && $313\pm 67$($335\pm 68$)  & $250(254)$ & $230\pm 12$& $272\pm 20$ & $229\pm20^{+31}_{-16}$& $251$\\
 $f_{B^*_s}^\perp$ && $297\pm 61$($302\pm58$)  &  & &
 & & \\
 $f_{B_c}$ && \underline{$360$}(\underline{$360$}) & $377(508)$ & $438\pm 10$ & $322\pm 42$ & &  \\
 $f_{B^*_c}$ && $387$($423$)  & $398(551)$ & $453\pm 20$ & $418\pm 24$ & & \\
 $f_{B^*_c}^\perp$ && $374$($392$)  &  & & &&\\\hline
 % $f_{\eta_b}$ & & $667\pm 4$ & $507(897)$ &  &  & &   \\
 %$f_{\Upsilon}$ &$715\pm 5$\cite{PDG08}& \underline{$715\pm 5$}  & $529(983)$ &  & $498\pm 20$ & & \\
 %$f_{\Upsilon}^\perp$ && $691\pm 5$  &  & &  & & \\\hline
 \end{tabular}}
%\end{ruledtabular}
 \end{table}
 \begin{table}%[ht!]
 \caption{\label{tab:decayconstantr} Ratio of the decay constants. In this work,
 $f^\perp_V/f_P=(1+f_V/f_P)/2$.} %The ratio $f_V/f_P$ is
%also included.
%\begin{ruledtabular}
 {\footnotesize
 \begin{tabular}{|c|c|c|c|c|c|c|c|}\hline
 & Experiment & This work\footnote{The value is obtained by $\varphi^g(\varphi^p)$.}
  & Linear(HO)  & FC\cite{FC}  & BS\cite{BS1,BS2} & Lattice\cite{Bec}
 & RQM\cite{Ebert}\\\hline
 $f_{D^*}/f_D$ & &$1.26\pm 0.02$ & $1.20(1.18)$ & $1.27\pm0.05$ & $1.48\pm 0.26$ &  &$1.32$  \\
 &&($1.49\pm0.02$)&&&&&\\%\hline
 $f_{D^*_s}/f_{D_s}$& & $ 1.27\pm 0.03$ & $1.17(1.15)$& $1.17\pm0.04$ &$1.51\pm 0.26$ & & $1.18$\\
 &&($1.51\pm0.03$)&&&&&\\%\hline
 $f_{D_s}/f_{D}$ &$1.27\pm 0.06$
 \cite{RS10} & $1.30\pm 0.04$ & $1.18(1.20)$ & $1.24\pm 0.03$ &
 $1.08\pm 0.01$& $1.10\pm 0.02$ &$1.15$ \\
 &&($1.26\pm0.04$)&&&&&\\%\hline
 $f_{D^*_s}/f_{D^*}$ & & $1.30\pm 0.05 $ & $1.14(1.18)$ &  &$1.10\pm 0.06$ & $1.11\pm 0.03$ & $1.02$\\
 &&($1.28\pm0.05$)&&&&&\\%\hline
 %$f_{J/\psi}/f_{\eta_c}$ & $1.24\pm 0.30$\footnote{This value is obtained by combining
 %$f_{J/\psi}=416\pm 7$ MeV \cite{PDG08} and
 %$f_{\eta_c}=335\pm 75$ MeV \cite{CLEO01}.} & $1.17\pm 0.00$ &$1.10(1.12)$ & & & & \\
 $f_{B^*}/f_B$ & &$1.10\pm 0.02$ & $1.08(1.07)$ & $1.08\pm0.04$ & $1.21\pm 0.27$ &
  &$1.16$  \\
   &&($1.22\pm0.03$)&&&&&\\%\hline
 $f_{B^*_s}/f_{B_s}$& & $1.11\pm 0.03$ & $1.07(1.07)$& $1.07\pm0.04$ &$1.26\pm 0.28$ & & $1.15$\\
 &&($1.24\pm0.05$)&&&&&\\%\hline
 $f_{B_s}/f_{B}$ & & $1.38\pm 0.07$ &  $1.24(1.32)$ &$1.19\pm 0.03$  & $1.10\pm 0.01$& $1.14\pm 0.03^{+0.00}_{-0.01}$ &$1.15$ \\
 &&($1.32\pm0.08$)&&&&&\\
 $f_{B^*_s}/f_{B^*}$ & & $1.39\pm 0.08$ & $1.23(1.32)$ &  & $1.14\pm 0.08$ & $1.17\pm 0.04^{+0.00}_{-0.03}$ &$1.15$ \\
 &&($1.35\pm0.08$)&&&&&\\
 $f_{B^*_c}/f_{B_c}$ && $1.08$($1.18$)  & $1.06(1.08)$ & $1.03\pm 0.03$ & $1.30\pm 0.24$ & &
 \\ \hline
 %$f_{\Upsilon}/f_{\eta_b}$ & & $1.07\pm 0.00$ & $1.04(1.10)$ & & & & \\ \hline
 \end{tabular}}
%\end{ruledtabular}
 \end{table}
\noindent In a previous work \cite{fpHwang}, we pointed out that
ratios $f_{D_s}/f_D$ and $f_{B_s}/f_B$ were not only chiefly
determined by the ratio of light quark masses, $m_s/m_q$, or the
$SU(3)$ symmetry breaking, but also insensitive to the heavy quark
masses $m_{c,b}$. This phenomenon also appears in the ratios
$f_{D^*_s}/f_{D^*}$ and $f_{B^*_s}/f_{B^*}$ here for both Gaussian
and power-law wave functions. On the contrary, as shown in Table 3,
the ratio $f_V/f_P$ is not only dependent on the heavy quark mass,
but also insensitive to the light quark mass. The reason is that,
making a comparison between Eqs. (\ref{Su}) and (\ref{V}), the
difference between $f_P$ and $f_V$ is proportional to $2
\kappa^2_\perp /(M_0+m_1+m2)$. In the Gaussian (power-law) wave
function, the mean square value of the transverse momentum is equal
(proportional) to the square value of the parameter $\beta$, or
$\langle \kappa^2_\perp \rangle= \beta^2(\langle \kappa^2_\perp
\rangle \propto \beta^2)$, so the ratio $f_V/f_P$ is influenced by
the parameter $\beta$ and the quark mass. In the case of the
different heavy quark, for example, $f_{D^*}/f_D$ and $f_{B^*}/f_B$,
as $m_b$ is much greater than $m_c$, this effect is greater than
that of $\beta_{bq} > \beta_{cq}$, so $f_{B^*}/f_B$ is smaller than
$f_{D^*}/f_D$. On the other hand, in the case of the different light
quark, for example, $f_{D^*}/f_D$ and $f_{D_s^*}/f_{D_s}$, as $m_s$
is slightly greater than $m_q$, this effect is less than that of
$\beta_{cs} > \beta_{cq}$, so $f_{D_s^*}/f_{D_s}$ is a little larger
than $f_{D^*}/f_D$.

The quark distributions of the heavy meson, $\phi_P(u)$,
$\phi_{V\|}(u)$, and $\phi_{V\perp}(u)$ are plotted in Fig. 1 and 2.
\begin{figure}
 \includegraphics*[width=4in]{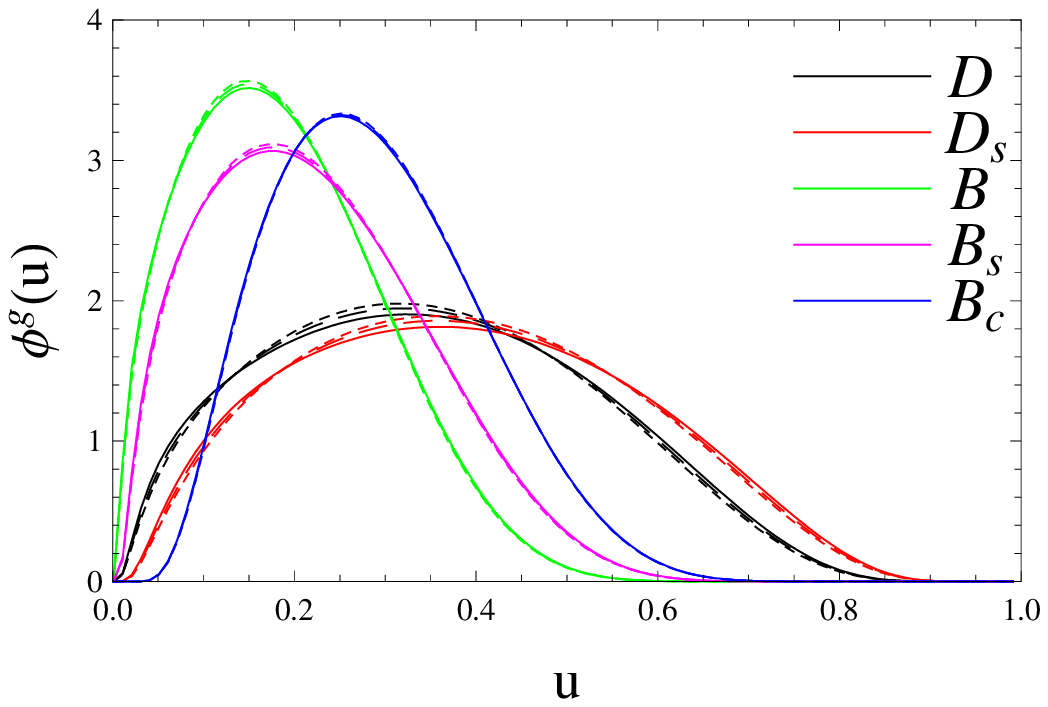}
 \caption{Quark distribution amplitudes of the heavy meson for a Gaussian wave function. The
 solid, dotted, and dashed lines correspond to $\phi^g_P(u)$,
 $\phi^g_{V\|}(u)$, $\phi^g_{V\perp}(u)$, respectively.
}
  \label{fig:phicb}
 \end{figure}
 \begin{figure}
 \includegraphics*[width=4in]{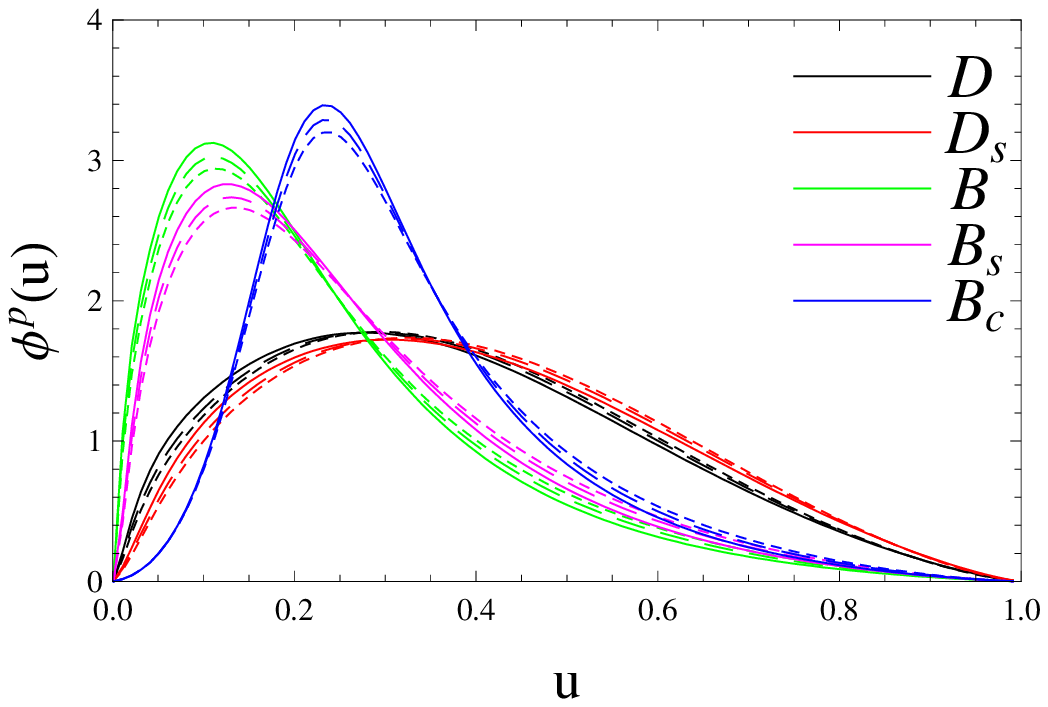}
 \caption{Quark distribution amplitudes of the heavy meson for a power-law wave function. The
 solid, dotted, and dashed lines correspond to $\phi^p_P(u)$,
 $\phi^p_{V\|}(u)$, $\phi^p_{V\perp}(u)$, respectively.
}
  \label{fig:phicbp}
 \end{figure}
Clearly %in the heavy quarkonium states, the major
%difference between the two curves of $\phi_{eta_c}(u)$ and
%$\phi_{\eta_b}(u)$ is that the latter, on which $u$ is peaked around
%zero, is sharper than the former. This means that the momentum
%fraction $u$ in $\eta_b$ is more centered on 1/2 than in $\eta_c$;
%moreover, in the heavy-light meson,
the difference in the constituent quark masses is greater, the
location where $u$ peaked is closer to zero. This indicates,
relatively, that the lighter the quark, the smaller its momentum
fraction. We also find that, even though the difference between
$f_D$ and $f_{D^*}$ was more than $25\%$ (almost $50\%$) for the
Gaussian (power-law) wave function, all curvilinear distinctions
between $\phi_{V\|}(u)$ and $\phi_P(u)$ are quite small. The reason
is that, after the $\kappa_\perp$ integration, the curve of
$\phi_M(u)$ is influenced only by the quark mass, parameter $\beta$,
and the total spin (that is, the pseudoscalar or the vector meson).
As the quark mass and $\beta$ are the same in
$\phi_{V\|(\perp)}(u)$ and $\phi_P(u)$, %and the normalization of
%$\phi_M(u)$,
the distinctions between them were slight. On the other hand, even
though $f_{D^*}$ is almost equal to $f_{D_s}$, as shown in Table 2
(or $f_{D^*}/f_D \simeq f_{D_s}/f_D$) for the Gaussian wave
function, the curvilinear distinction between $\phi^g_{D^*\|}(u)$
and $\phi^g_{D}(u)$ is obviously smaller than that between
$\phi^g_{D_s}(u)$ and $\phi^g_{D}(u)$. As for the power-law wave
function, the situation is inverse. Therefore, we can infer that,
even though the values of $f_M$'s are almost the same between the
distinct heavy mesons, the curves of $\phi_M(u)$ may have the quite
large differences, and vice versa. Finally, the quark distribution
function is displayed in terms of the $\xi$-moments, as in Eq.
(\ref{ximoments}). The first six $\xi$-moments ($n
> 0$) are listed in Table 4.

\begin{table}[ht!]
\caption{\label{tab:cximoment} First six $\xi$-moments of the
$s$-wave heavy meson. }
%\begin{ruledtabular}
{\footnotesize
 \begin{tabular}{|c|c|c|c|c|c|c|}\hline
  & $\langle \xi^1\rangle$ & $\langle \xi^2\rangle$ & $\langle \xi^3\rangle$
 & $\langle \xi^4\rangle$& $\langle \xi^5\rangle$& $\langle \xi^6\rangle$\\\hline
 $\phi^{g(p)}_D$ & $-0.288$($-0.251$)  & $0.210$($0.235$) & $-0.125$($-0.115$) &
 $0.0960$($0.111$) & $-0.0695$($-0.0673$) & $0.0558$($0.0664$) \\
 $\phi^{g(p)}_{D_s}$ & $-0.213$($-0.207$) & $0.183$($0.217$) & $-0.0890$($-0.0905$)
 & $0.0738$($0.0970$) & $-0.0468$($-0.0507$) & $0.0388$($0.0550$)\\
 $\phi^{g(p)}_B$ & $-0.617$($-0.531$)& $0.425$($0.398$) & $-0.312$($-0.288$)
 &$0.240$($0.234$) & $-0.191$($-0.185$) & $0.156$($0.157$)\\
 $\phi^{g(p)}_{B_s}$ & $-0.549$($-0.486$)  &$0.359$($0.359$)& $-0.254$($-0.249$)
 &$0.189$($0.200$) & $-0.147$($-0.154$)& $0.117$($0.129$)\\
 $\phi^{g(p)}_{B_c}$ & $-0.536$($-0.368$) &$0.227$($0.230$)&$-0.133$($-0.123$)&
 $0.108$($0.0867$)& $-0.0553$($-0.0527$)& $0.0378$($0.0403$)\\ \hline
 %$\xi$-moments & $\langle \xi^2\rangle$ & $\langle \xi^4\rangle$ & $\langle \xi^6\rangle$ \\ \hline
 %$\phi_{\eta_c}$ & $0.111$  & $0.0280$  & $0.00959$ \\
 %$\phi_{\eta_b}$ & $0.0573$  & $0.00830$ & $0.00175$\\ \hline
\end{tabular}}
%\end{ruledtabular}
\end{table}

For the heavy quark framework, some models for B meson LCDAs have
also been adopted in the literature. Inspired by the QCD sum rule
analysis, the authors of Ref. \cite{GN} proposed a simple model:
 \be \label{GNphi}
 \psi_{+I}(\omega)=\frac{\omega}{\lambda_I^2}
 e^{-\omega/\lambda_I}.
 \en
Additionally, the authors of Ref. \cite{HWZ} suggested a
Gaussian-type model:
 \be \label{HWzphi}
 \psi_{+II}(\omega)=\sqrt{\frac{2}{\pi \lambda^2_{II}}}\frac{\omega^2}{\lambda^2_{II}}
 e^{-\omega^2/2\lambda^2_{II}}.
 \en
By applying Eqs. (\ref{momentw}) and (\ref{omegaLambda}), the
relation between the residual center mass and the parameter
$\lambda$ is:
 \be
 \bar \Lambda_q = \frac{3}{2} \lambda_{I}=
 \frac{3}{\sqrt{2\pi}}\lambda_{II}.
 \en
In Ref. \cite{LN}, the value $\lambda_I=0.3$ GeV corresponded to
$\bar \Lambda_q =0.45$ GeV. For a convenient comparison, we used
this $\bar \Lambda_q$ and $\bar \Lambda_s=\bar \Lambda_q+m_s-m_q$ to
fix parameters $\beta_{Qq}$ and $\beta_{Qs}$ in this work. Moreover,
the Gaussian wave function $\hat{\varphi}$ is given by taking the
heavy quark limit in Eq. (\ref{Gaussian1s}) % and (\ref{Powerlaw1s})
and using the relation Eq. (\ref{scale}):
 \be
 \hat{\varphi}(\omega,\kappa_\perp)=4\bigg(\frac{\pi}
 {\beta^2}\bigg)^{3/4}\sqrt{\frac{1}{2}+\frac{m^2_2+\kappa^2_\perp}{2
 \omega^2}}~{\rm exp}
 \bigg(-\frac{\kappa_\perp^2+(\frac{\omega}{2}
 -\frac{m^2_2+\kappa^2_\perp}{2
 \omega})^2}{2\beta^2}\bigg).\label{Gaussian1sH}%\\
 %\hat{\varphi}^p(\omega,\kappa_\perp)&=&8\bigg(\frac{2\pi}
 %{\beta^3}\bigg)^{1/2}\sqrt{\frac{1}{2}+\frac{m^2_2+\kappa^2_\perp}{2
 %\omega^2}}~
 %\Bigg[\frac{\beta^2}{\kappa_\perp^2+(\frac{\omega}{2}
 %-\frac{m^2_2+\kappa^2_\perp}{2 \omega})^2+\beta^2}\Bigg]^2.\label{Powerlaw1sH}
 \en
The light quark masses $m_{q(s)}=0.251(0.445)$ GeV are as in Table
1, and we can then obtain the values $\beta^g_{Qq}=0.279$ GeV and
$\beta^g_{Qs}=0.338$ GeV.
%$\beta^p_{Qq}=0.0186$ GeV, and $\beta^p_{Qs}=0.0194$ GeV.
%These values are listed in Table V.
%\begin{table}[ht!]
%\caption{\label{tab:parameterH} Decay constants (GeV$^{3/2}$) and
%$\beta$'s (GeV) for the heavy quark framework. }
%%\begin{ruledtabular}
% {\footnotesize
%\begin{tabular}{|c|c|c|c|}\hline
% $F_{Qq}$ & $F_{Qs}$ & $\beta_{Qq}$  & $\beta_{Qs}$  \\ \hline
% $0.469\pm 0.071$ & $0.651\pm 0.125$ & $0.536\pm 0.058$ & $0.650\pm 0.090$ \\ \hline
%\end{tabular}}
%\end{ruledtabular}
%\end{table}
In terms of these parameters, the leading twist LCDAs
$\Phi_{Qq}(\omega)$, $\Phi_{Qs}(\omega)$, $\psi_{+I}(\omega)$, and
$\psi_{+II}(\omega)$ are calculated and plotted as in Fig. 3.
\begin{figure}
 \includegraphics*[width=4in]{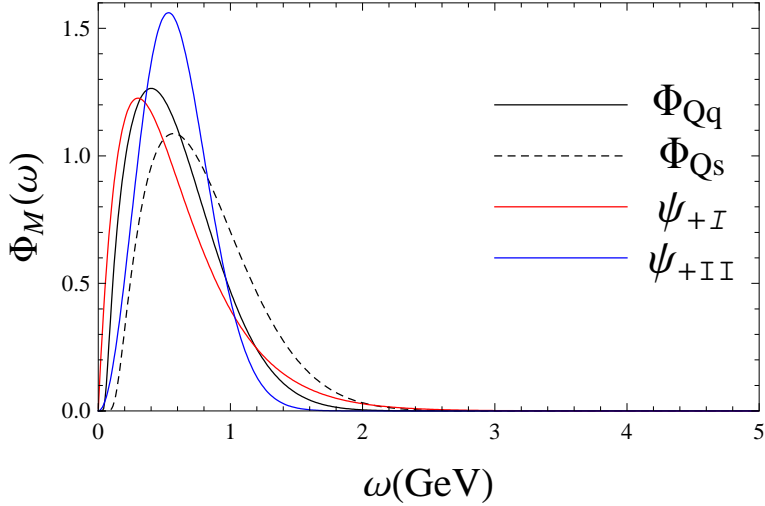}
 \caption{Leading twist LCDAs in the heavy quark framework. %The
  %black solid and dotted lines correspond to $\Phi_{Qq}(\omega)$ and
 %$\Phi_{Qs}(\omega)$, respectively. The red and blue lines correspond to $\psi_{+I}(\omega)$ and
 %$\psi_{+II}(\omega)$, respectively. %The forms of
 %$\psi_{+I}(\omega)$ and $\psi_{+II}(\omega)$ are given in Eqs.
 %(\ref{GNphi}) \cite{GN,LN} and (\ref{HWzphi}) \cite{HWZ}, respectively.
 }
  \label{fig:Phi21}
 \end{figure}
We find that the curve of $\Phi_{Qq}(\omega)$ is close to that of
$\psi_{+I}(\omega)$.
%The fist moments $\langle \omega \rangle$ of $\Phi_{Qq}(\omega)$ and
%$\Phi_{Qs}(\omega)$ are obtained as $\langle \omega
%\rangle_{Qq}=1.09$ GeV and $\langle \omega \rangle_{Qs}=1.35$ GeV,
%respectively. From Eq. (\ref{omegaLambda}), the corresponding
%residual center masses are $\bar \Lambda_q =0.820$ GeV and $\bar
%\Lambda_s =1.01$ GeV. The residual center mass difference $\bar
%\Lambda_s-\bar \Lambda_q = 192$ MeV is consistent with the light
%quark mass difference $m_s-m_q = 194$ MeV.
%%%%%%%%%%%%%%%%%%%%%%
\section{Conclusions}
This study has discussed the leading twist LCDAs of the $s$-wave
heavy meson within the light-front approach in both general and
heavy quark frameworks. These LCDAs are shown in terms of
light-front variables and relevant decay constants. In the general
frameworks, we find that the decay constants and LCDAs of the
pseudoscalar and vector mesons have the following relations:
$f_P+f_{V}=2 f_V^\perp$ and $\phi_P+\phi_{V\|}=2\phi_{V\perp}$. The
parameters $m$ and $\beta$, which appear in both Gaussian and
power-law wave functions, were determined as follows: (1) the light
quark masses are fitted by the decay constants and the mean square
radii of the light meson; (2) the heavy quark masses are determined
by the mass of the
spin-weighted average of the %triplet $p$-wave
heavy quarkonium states and its variational principle for the
relevant Hamiltonian; and (3) the hadronic parameter $\beta$'s are
evaluated by the decay constants of $D^+$, $B^-$, and $B_c$, with
the former two and the latter one from the experimental data and the
average result of QCD sum rules, respectively. We find that, for
both Gaussian and power-law wave functions, the ratios
$f_{D^*_s}/f_{D^*}$ and $f_{B^*_s}/f_{B^*}$, as well as
$f_{D_s}/f_D$ and $f_{B_s}/f_B$ in the previous work, are chiefly
determined by the ratio of light quark masses $m_s/m_q$, or the
$SU(3)$ symmetry breaking. On the other hand, by making a comparison
between $f_{D^*}/f_D$, $f_{D_s^*}/f_{D_s}$, $f_{B^*}/f_B$, and
$f_{B_s^*}/f_{B_s}$, the ratio $f_V/f_P$ is not only dependent on
the heavy quark mass, but also insensitive to the light quark mass.

As shown in Fig.1 and 2, we find that even though the difference
between $f_D$ and $f_{D^*}$ is more than $25\%$ (almost $50\%$) for
the Gaussian (power-law) wave functions, all curvilinear
distinctions between $\phi_{V\|}(u)$ and $\phi_P(u)$ are quite small
because their main difference come from the variations of the quark
mass and $\beta$. On the contrary, even though $f_{D^*}/f_D$ is
almost equal to $f_{D_s}/f_D$ for the Gaussian wave function, the
curvilinear distinction between $\phi^g_{D^*\|}(u)$ and
$\phi^g_{D}(u)$ is obviously smaller than that between
$\phi^g_{D_s}(u)$ and $\phi^g_{D}(u)$. As for the power-law wave
function, the situation is inverse. Therefore, we conclude that even
though the values of $f_M$'s are almost equal among the distinct
mesons, the curves of $\phi_M(u)$ may have quite large differences,
and vice versa.

When the heavy quark framework is used, the above relations for the
decay constant and the LCDAs can be further simplified as $F_P =
F_{V} = F_V^\perp$ and $\Phi_P= \Phi_{V\|}= \Phi_{V\perp}$, as
consistent with HQS. For a convenient comparison, the value $\bar
\Lambda_q=0.45$ GeV, as suggested in Ref. \cite{LN}, is used to fix
$\beta_{Qq(Qs)}$ and to plot the curves of $\Phi_{Qq}$, $\Phi_{Qs}$,
$\psi_{+I}$, and $\psi_{+II}$ in Fig. 3. We find that the
curvilinear distinction between $\Phi_{Qq}(\omega)$ and
$\psi_{+I}(\omega)$ is relatively small.

%%%%%%%%%%%%%%%%%%%%%%%%%%%%%%%%%%%%%%%%%%%%%%%%%%%%%%%%%%%%%%%%%%%%%%%%%

{\bf Acknowledgements}\\
This work was supported in part by the National Science Council of
R.O.C. under Grant No NSC-96-2112-M-017-002-MY3.

%%%%%%%%%%%%%%%%%%%%%%%%%%%%%%%%%%%%%%%%%%%%%%%%%%%%%%%%%%%%%%%%%%%%%%%%%%%%%%%%%%%%%%%%

\end{document}